\documentclass[reprint,aps,prc,nofootinbib]{revtex4-1}

\usepackage{amsmath}
\usepackage{amsfonts}
\usepackage{amssymb}
\usepackage{physics}
\usepackage{slashed}
\usepackage{simplewick}

\usepackage{graphicx}
\usepackage[caption=false]{subfig}

\usepackage{listings}

\usepackage{tikz}
\usetikzlibrary{calc}
\usetikzlibrary{patterns}
\usetikzlibrary{patterns}
\usetikzlibrary{positioning,decorations.pathmorphing,decorations.markings}

\usepackage{soul}
\usepackage[utf8]{inputenc}

\usepackage{feynmp-auto}
\usepackage{multirow}

\usepackage{xcolor}

\usepackage{graphicx}
\usepackage{float}
\usepackage{overpic}

\usepackage{hyperref}

\begin{document}

\title{An Improved Hadronic Model for Pion Electroproduction}

\author{Robert J. Perry}
\email{perryrobertjames@gmail.com}
\affiliation{Institute of Physics, National Chiao-Tung University, 1001 Ta-Hsueh Road, Hsinchu 30010, Taiwan}
\author{Ay{\c s}e K{\i}z{\i}lers{\" u} }%
\author{Anthony W. Thomas}
\affiliation{CSSM and ARC Centre of Excellence for Particle Physics at the Tera-scale, Department of Physics, University of Adelaide, Adelaide SA 5005, Australia}%

\date{\today}

\begin{abstract}
Current measurements of the high energy behavior of the pion form factor are obtained from pion electroproduction data. These values are model dependent, utilizing the Vanderhaeghen, Guidal and Laget Regge (VGL) Model for their extraction. Recent work which examined the implementation of gauge invariance in that model suggested that it might lead to extracted pion form factors larger than the true values. Here we introduce a new model which preserves the successes of the VGL Model but implements gauge invariance in a new way. To demonstrate the validity of this new approach, we first use it to extract the pion form factor in a simple toy model. When compared with the previous extraction method, the improved model leads to a more reliable extraction. The success in this simple model leads us to reanalyse the electroproduction cross section data, where we obtain comparable values for the pion form factor to those obtained using the VGL procedure.

\end{abstract}

\pacs{Valid PACS appear here}
\maketitle

\section{Introduction}

\begin{figure}
\centering
\includegraphics[scale=0.4]{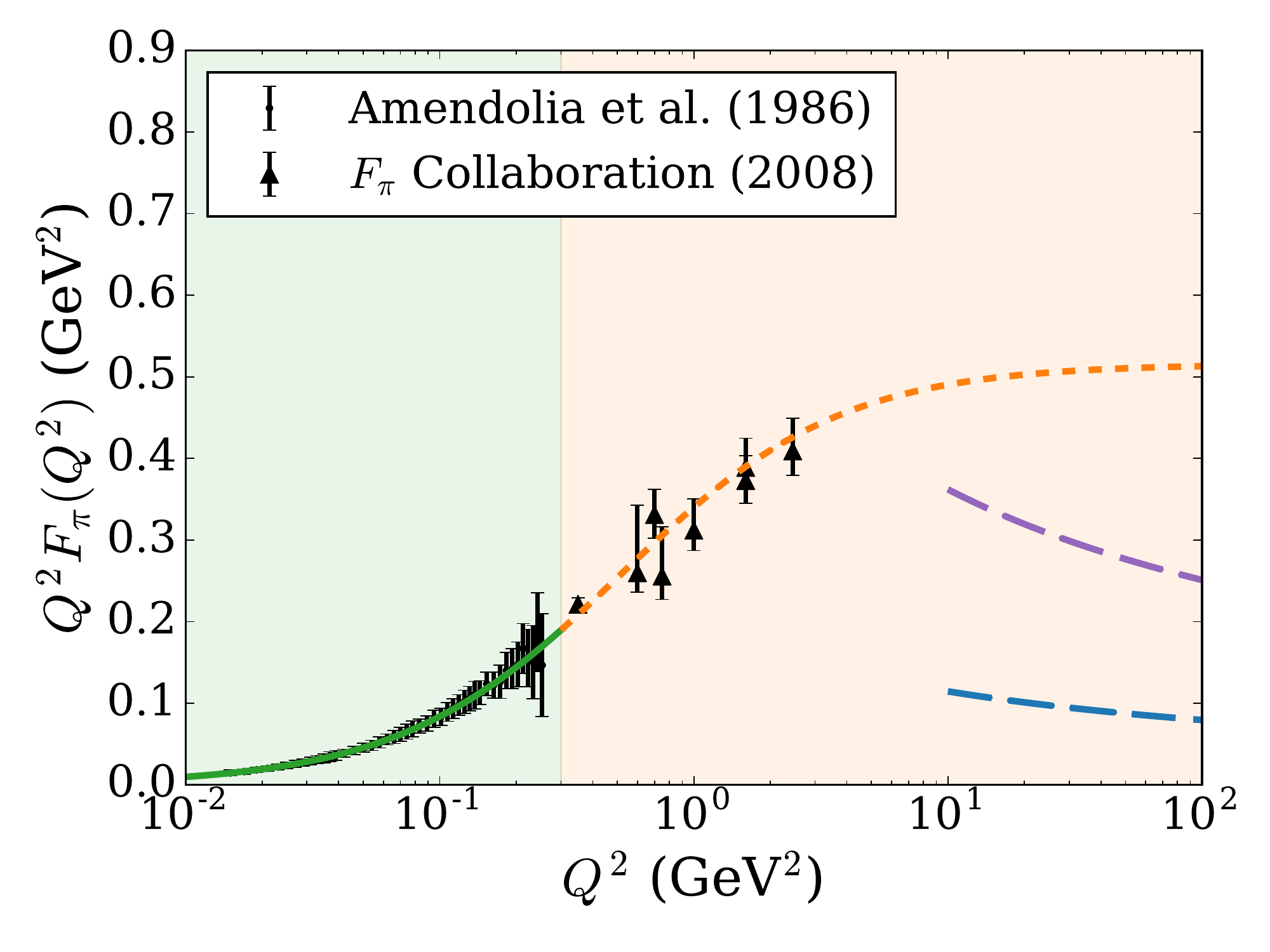}
\caption{Current measurements of the pion's electromagnetic form factor $F_\pi(Q^2)$, multiplied by the photon virtuality. The upper dashed curve constitutes a monopole parameterization based on fitting the low energy data to a monopole shape, as predicted by vector meson dominance (VMD) models. Two versions of the prediction of Lepage and Brodsky are shown: the lowest curve corresponds to the true asymptotic prediction given in  Eq.~ref{eq:lepageBrodsky}, while the central curve is the result of incorporating a non-asymptotic form of the pion's light cone distribution amplitude utilizing the results of Ref~\cite{Chang:2013nia}.}\label{fig:1}
\end{figure}

The pion's electromagnetic form factor is of theoretical interest in part because of the work of Lepage and Brodsky which predicted that at very large photon virtualities the pion form factor should scale as~\cite{Lepage:1979zb}:
\begin{equation}\label{eq:lepageBrodsky}
Q^2F_\pi(Q^2)\to 16\pi f_\pi^2\alpha_s(Q^2) \, ,
\end{equation}
where $f_\pi\approx0.132$ GeV is the pion decay constant, and $\alpha_S(Q^2)$ is the strong coupling constant. Naively, one might hope that the transition to this behavior would occur around the QCD mass scale $\Lambda_\text{QCD}\sim0.2$ GeV. However, as we can see from the current measurements of the pion form factor shown in Fig.~\ref{fig:1}, this does not seem to be the case. Instead, the pion form factor appears to follow the monopole form predicted from vector meson dominance for all kinematic points measured.

In fact, the predicted behavior of the pion form factor described in Eq.~\ref{eq:lepageBrodsky} is actually a simplification of the relation derived by Lepage and Brodsky which is given as
\begin{equation}
\begin{split}
Q^2F_\pi(Q^2)&
\\
=16\pi f_\pi^2&\alpha_s(Q^2/\mu^2)\omega_\phi^2(Q^2/\mu^2)+\mathcal{O}(\alpha_s(Q^2/\mu^2)^2)
\end{split}
\end{equation}
where the additional factor $\omega_\phi$ is related to the pion’s valence quark parton distribution amplitude $\phi_\pi(x,Q^2/\mu^2)$, and is given as
\begin{equation}
\omega_\phi(Q^2/\mu^2)=\frac{1}{3}\int_0^1 dx\frac{1}{x}\phi_\pi(x,Q^2/\mu^2) \, .
\end{equation}
One may write the general solution for the pion’s valence quark parton distribution amplitude as
\begin{equation}
\begin{split}
\phi_\pi(x,Q^2/\mu^2)&
\\
=x(1-x)&\sum_{n=0,2,\dots}^\infty a_nC_n^{3/2}(2x+1)\bigg(\ln\frac{Q^2}{\mu^2}\bigg)^{-\gamma_n}
\end{split}
\end{equation}
where $C_n^{3/2}(x)$ are the Gegenbauer polynomials and $\gamma_n\geq0$ are the anomalous dimensions. For photon virtualities far above the renormalization scale, the subleading terms are suppressed and in this limit one finds
\begin{equation}
\phi_\pi(x,Q^2/\mu^2)\to x(1-x)a_0C_0^{3/2}(2x+1)=a_0x(1-x) \, ,
\end{equation}
where $a_0=6$ is fixed by a sum rule~\cite{Lepage:1979zb}. In this limit, $\omega_\phi(Q^2/\mu^2)=1$, and the more well known relation is obtained. Modern predictions of this asymptotic behavior which are arguably more suitable for the energies probed in today's experiments have reduced the discrepancy between the experimental data and the theoretical prediction~\cite{Chang:2013nia}. However, the theory still appears to underestimate the form factor~\cite{Chang:2013nia}. It is therefore of interest to examine other possible reasons for this difference. 

Currently, the high virtuality data for the pion form factor is obtained from pion electroproduction. This approach has previously been criticized for a number of reasons.

Firstly, others have argued that the object truly measured in pion electroproduction is the transition amplitude between a mesonic state with an effective space-like mass $t<0$ and the physical pion~\cite{Brodsky:1997dh}. It has been argued that this transition amplitude may be  larger than the physical pion form factor~\cite{Brodsky:1997dh}.

Secondly, there have been questions raised about the particular theoretical model known as the Vanderhaeghen, Guidal and Laget Regge Model (VGL Model)~\cite{Vanderhaeghen:1997ts,Guidal:1997by} utilized in the analysis~\cite{Perry:2018kok}. In particular, the method by which gauge invariance is imposed on the amplitude appears unnatural. The method amounts to requiring that the pion's electromagnetic form factor $F_\pi(Q^2)$ and the proton's Dirac form factor $F_1^p(Q^2)$ are equal. Recently, the extraction method used in the most recent measurement was applied to a toy model of electroproduction to further investigate the efficacy of the approach. It was found that the VGL Model led to the overextraction of the toy form factor, suggesting that the measured form factor data may be overestimated~\cite{Perry:2018kok}. This study utilized a rather simple model of electroproduction and thus questions may be raised about the size of any observed effects. However, the trend towards less accurate measurements of the pion form factor at higher $Q^2$ is important to understand, especially with the planned new set of measurements at Jefferson Laboratory of the pion's electromagnetic form factor at higher photon virtuality than ever before~\cite{Dudek:2012vr}.

In this paper, we propose a modified version of the VGL Model, which we term the Gauge Improved VGL Model. By modifying the way gauge invariance is imposed on the amplitude, we no longer need to require that the pion and proton form factors are equal. To do this, we begin by introducing the conventions followed throughout this paper. We then explain our modified VGL Model, before reanalyzing the experimental data.

\section{Kinematics and Preliminaries}

We focus on describing the reduced $2\to2$ scattering amplitude $p(p_1)+\gamma^*(q)\to n(p_2)+\pi(p_\pi) $. We introduce conventional Mandelstam variables for this process, and we define $Q^2=-q^2$ so that the photon's spacelike momenta is positive. These three momenta ($Q^2$, the proton-photon invariant mass $W=\sqrt{s}$ and $t$) allow one to fully describe the cross section. The unpolarized differential cross section may be separated according to the polarization states of the virtual photon into transverse ($T$), longitudinal ($L$) polarizations, as well as two interference terms ($LT$ and $TT$)~\cite{Blok:2008jy}:
\begin{equation}
\begin{split}
(2\pi)\frac{d^2\sigma}{dtd\phi}\,\,=\,\,&\frac{d\sigma_T}{dt}+\epsilon\frac{d\sigma_L}{dt}  
\\
+&\sqrt{2\epsilon(\epsilon+1)}\frac{d\sigma_{LT}}{dt}\cos\phi+\epsilon\frac{d\sigma_{TT}}{dt}\cos2\phi  ,
\end{split}
\end{equation}
where $\epsilon$ is a measure of the virtual photon polarization~\cite{Blok:2008jy,Huber:2008id}. The $t$-channel pion exchange diagram dominates the longitudinal differential cross section $d\sigma_L/dt$~\cite{Kaskulov:2008xc}. It is this structure function which we aim to describe effectively. Details on the relationship between the invariant matrix element $i\mathcal{M}^\mu$ we derive for this process and the cross section can be found in Ref.~\cite{Amaldi:1979vh}.


\section{The Born Term Model and the VGL Model}

\begin{figure}
\centering
\includegraphics[scale=1]{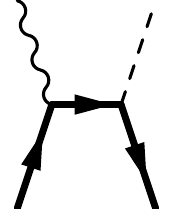} 
\includegraphics[scale=1]{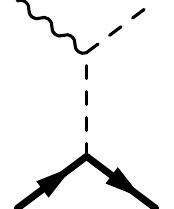} 
\caption{Born Term Model for pion electro-production. The pion form factor is measured in pion electroproduction via the $t$-channel diagram. There is no $u$ channel diagram because in our effective field theory, the neutron is neutral at tree level.}\label{fig:BTM}
\end{figure}

Either the pseudo-vector or pseudo-scalar realizations of pion-nucleon effective field theory may be used since for this process both may be shown to give the same matrix element. More discussion of these Lagrangians and their corresponding Feynman rules may be found in Ref.~\cite{Ji:2013bca}. The Born Term Model is defined by the tree level diagrams, which are shown in Fig.~\ref{fig:BTM}. 
%

The VGL Model is a Regge Model. One may understand the Reggeization of the amplitude as the multiplication of the Born Term Model by the ratio of the reggeized propagator to the Born Term Model propagator. The structure of the pion is further incorporated by multiplying this amplitude by an overall factor of the pion form factor $F_\pi(Q^2)$. That is
\begin{equation}
i\mathcal{M}_{\text{VGL}}^\mu=F_\pi(Q^2) D_F^{\pi-1}(p_t)D_R^\pi(p_t)\big[i\mathcal{M}_{\text{BTM}}^\mu\big].
\end{equation}

Introducing these terms as overall mutiplicative factors is motivated by gauge invariance arguments~\cite{Vanderhaeghen:1997ts,Perry:2018kok}. This `factorization' of the pion form factor is rather unnatural. One may view this as a model assumption that the pion and proton form factors are equal:
\begin{equation}
F_\pi(Q^2)\underset{\text{VGL}}{\approx} F_1^p(Q^2) \, .
\end{equation}
Of course, at the pion pole this introduces no error, however the data is some distance from the pole. The purpose of this paper is to discuss a method by which we may implement structure at the pion electromagnetic vertex in a way consistent with gauge invariance, without being required to modify the proton's electromagnetic vertex.


\section{The Gauge Improved VGL Model}

\subsection{Pion Electroproduction Vertex}

In pion electroproduction the most general form of the pion-photon vertex will take the form 
\begin{equation}\label{eq:generalForm}
\begin{split}
\Gamma_\pi^\mu(k_1,k_2;q)&=(k_1+k_2)^\mu f_1(t,m_\pi^2;q^2)
\\
&+(k_1-k_2)^\mu f_2(t,m_\pi^2;q^2).
\end{split}
\end{equation}
We may use the Ward-Green-Takahashi Identity 
\begin{equation}
-iq_\mu \Gamma_\pi^\mu(k_1,k_2;q)=D_F^{-1}(k_2^2)-D_F^{-1}(k_1^2),
\end{equation}
where $D_F(k^2)$ is the most general form of the propagator;
\begin{equation} 
D_F(k^2)=\frac{i}{k^2-m^2-\Sigma(k^2)},
\end{equation}
and we identify $\Sigma(k^2)$ as the particle's renormalized self energy. We have
\begin{equation}
-[t-m_\pi^2-\Sigma(t)]=(m_\pi^2-t) f_1(t,m_\pi^2;q^2)-q^2 f_2(t,m_\pi^2;q^2).
\end{equation}
We can use this relation to constrain $f_2$:
\begin{equation}
\label{eq:GI}
f_2(t,m_\pi^2;q^2)=\frac{[t-m_\pi^2-\Sigma(t)]-(t-m_\pi^2) f_1(t,m_\pi^2;q^2)}{q^2}.
\end{equation}
The implication of this relation is that we are free to replace the Born Term Model electromagnetic vertex with the most general form of the electromagnetic vertex, \textit{provided} we relate the two form factors as in \eqref{eq:GI}. 

\subsection{The Gauge Improved VGL Model}

Here we define the Gauge Improved VGL Model. We begin with the Born Term Model, which may be written
\begin{equation}\label{eq:BTM}
\begin{split}
i\mathcal{M}_{\text{BTM}}^\mu&=\frac{g_A}{\sqrt{2}f_\pi}2m_N\overline{u}_N(p_2,\lambda_2)\gamma_5\bigg[S_F^N(p_1+q)(-ie\gamma^\mu)
\\
&+D_F^\pi(p_\pi-q)(-ie)\Gamma_{\pi0}^\mu\bigg]u_N(p_1,\lambda_1),
\end{split}
\end{equation}
where $\Gamma_{\pi0}^\mu=(p_t+p_\pi)^\mu$ is the pion's electromagnetic current, suitable at low energies, and $S_F^N$ and $D_F^\pi$ are the nucleon and pion propagators, respectively. Following our discussion in the last section, we replace this vertex function with the most general form (Eq.~\ref{eq:generalForm}), where $f_2$ is related to $f_1$ via Eq.~\ref{eq:GI}. We also multiply the matrix element by an overall form factor, which we take to be the physical proton Dirac form factor, $F_1^p(q^2)$. Thus our model is
\begin{equation}
\begin{split}
i\mathcal{M}_{\text{GIVGL}}^\mu&=\frac{g_Ae}{\sqrt{2}f_\pi}2m_NF_1^p(q^2)\overline{u}_N(p_2,\lambda_2)\gamma_5
\\
&\times\bigg[\frac{(\slashed{p}_1+\slashed{q}+m_N)}{s-m_N^2}\gamma^\mu+\frac{\Gamma_{\pi}^\mu}{t-m_\pi^2} \bigg]u_N(p_1,\lambda_1)
\\
&=\frac{g_Ae}{\sqrt{2}f_\pi}2m_N\overline{u}_N(p_2,\lambda_2)\gamma_5
\\
&\times\bigg[\frac{(\slashed{p}_1+\slashed{q}+m_N)}{s-m_N^2}F_1^p(q^2)\gamma^\mu+\frac{F_1^p(q^2)\Gamma_{\pi}^\mu}{t-m_\pi^2} \bigg]
\\
&\times u_N(p_1,\lambda_1)
.
\end{split}
\end{equation}
We have chosen to not Reggeize the model at this stage. We did check that in the later analysis of the experimental data Reggeizing the model made no significant difference to the final values of the extracted pion form factor. The Gauge Improved VGL Model has a single unconstrained function $f_1(t,m_\pi^2,q^2)$. In order to extract the pion form factor, we must now relate $f_1(t,m_\pi^2;q^2)$ to the physical pion form factor. Since this measurement is performed at small $t$, we perform a Taylor expansion around $t=m_\pi^2$.
\begin{equation}
\begin{split}
f_1(t,m_\pi^2;q^2)&=f_1(m_\pi^2,m_\pi^2;q^2)
\\
&+(t-m_\pi^2)\frac{d}{dt}f_1(t,m_\pi^2;q^2)\bigg|_{t=m_\pi^2}+\dots.
\end{split}
\end{equation}
The residue of the pion pole must be proportional to the pion form factor. Thus we associate 
\begin{equation}
F_\pi(q^2)=F_1^p(q^2)f_1(m_\pi^2,m_\pi^2;q^2).
\end{equation}
We further define 
\begin{equation}
g_1(q^2)=\frac{d}{dt}f_1(t,m_\pi^2;q^2)\bigg|_{t=m_\pi^2}.
\end{equation}
Thus we write
\begin{equation}
\begin{split}
f_1(t,m_\pi^2;q^2)=f_1(m_\pi^2,m_\pi^2;q^2)\bigg[1+(t-m_\pi^2)R(q^2)\bigg],
\end{split}
\end{equation}
where we note that the second term will produce an effective contact interaction, since the factor of $t-m_\pi^2$ will cancel the $t$-channel pole. We have defined
\begin{equation}
R(q^2)=\frac{g_1(q^2)}{f_1(m_\pi^2,m_\pi^2;q^2)}.
\end{equation}
The Gauge Improved VGL Model contains two free parameters ($f_1(q^2)$ and $g_1(q^2)$) rather than the one in the original VGL Model. As we shall see, this model is able to successfully describe the cross section data, with accuracy comparable to that of the VGL Model.

\section{Results}

\subsection{A Simple Toy Model}

\begin{figure}
\includegraphics[scale=0.4]{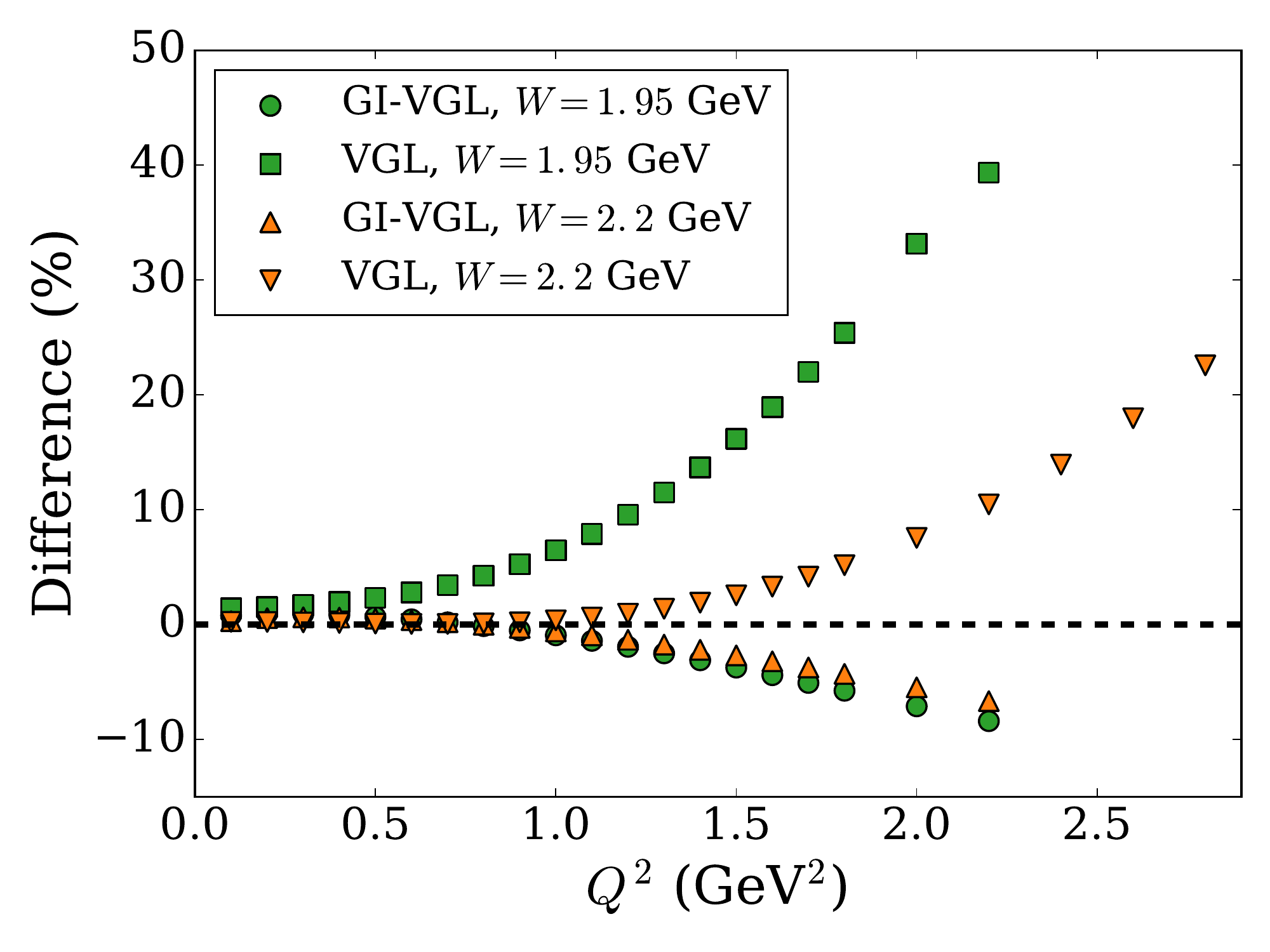}
\caption{Comparison of the difference between the true toy model form factor and the extracted value. Note that when compared with the previous VGL Model, our Gauge Improved VGL Model (denoted GI-VGL in the plot) achieves a better agreement with the true form factor for all values of $Q^2$ examined. Further, the Gauge Improved VGL Model has a much reduced sensitivity to the invariant mass $W$.}\label{fig:percentDiff}
\end{figure}

As a first study of the effectiveness of our model, we attempt to extract the pion form factor from the toy model of electroproduction described in Ref.~\cite{Perry:2018kok}. The results of this analysis are shown in Fig.~\ref{fig:percentDiff}. Using the Gauge Improved VGL Model leads to a measurement of the pion form factor which is less dependent on $W$, and also appears to give a better extraction of the true form factor at all kinematics studied. 

\begin{figure}
\includegraphics[scale=0.4]{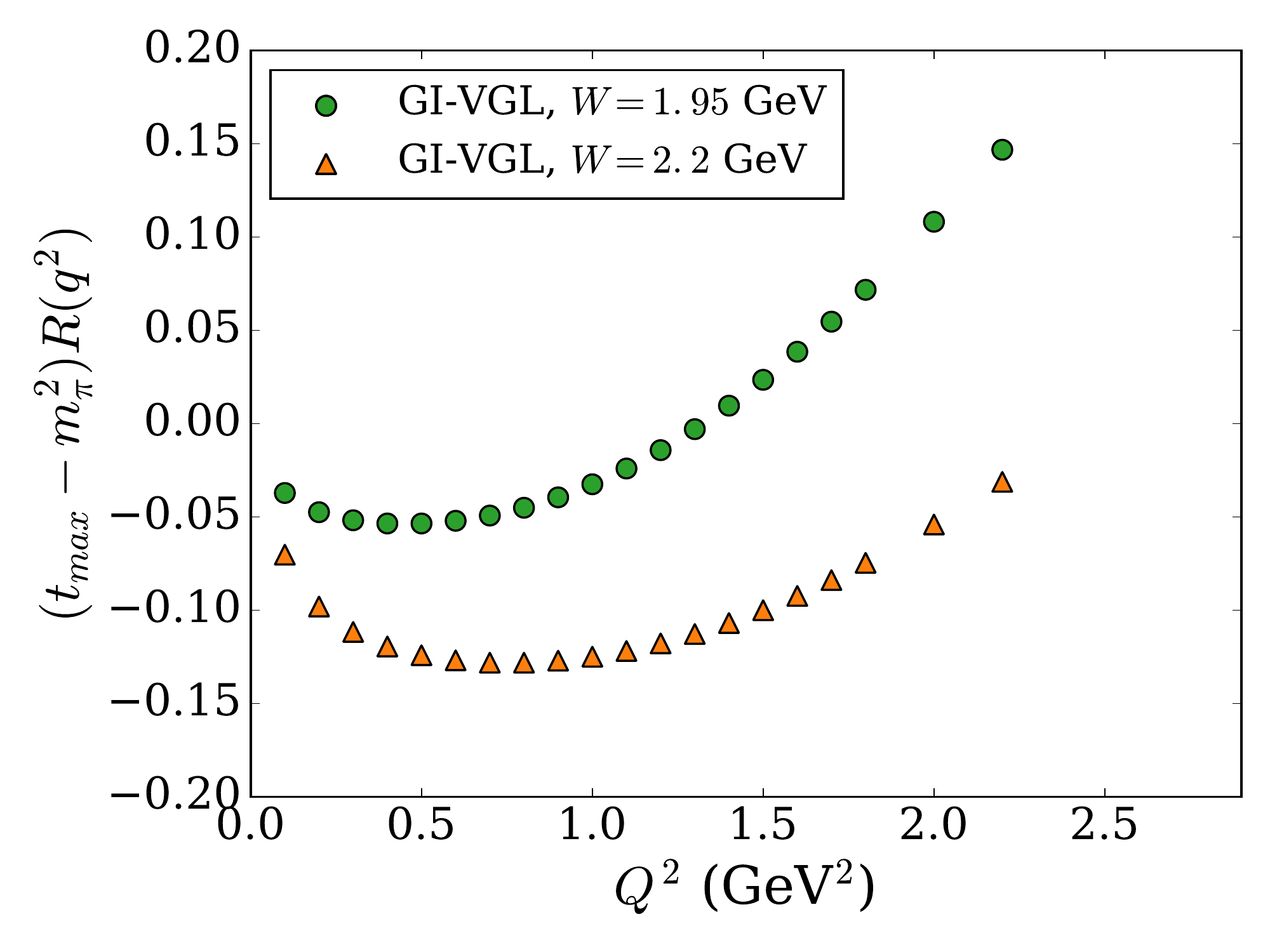}
\caption{Comparison of the magnitude of the first term in the Taylor series expansion of $f_1(t,m_\pi^2,q^2)$ compared to the size of the zeroth term. $t_{max}$ is the largest value of $-t$ for which we fit the model. Since this term increases linearly with $t$, this choice allows us to study the largest non-pole contribution.
}\label{fig:ratio_toyModel}
\end{figure}

We may also study the magnitude of the residue of the pole term when compared to the residue of the contact term, shown in Fig.~\ref{fig:ratio_toyModel}. We see that as the photon virtuality increases, so too does the contribution from the non-pole term. This occurs because for fixed $W$, the extrapolation distance in $t$ increases as $Q^2$ increases~\cite{Perry:2018kok}.

\subsection{Re-analysis of the Pion Form Factor From the Most Recent Electroproduction Data}

Having demonstrated the improvements of the Gauge Improved VGL Model, we proceed to re-analyze the most recent electroproduction data~\cite{Huber:2008id,Blok:2008jy}.

\begin{figure}[h]
\centering
\subfloat[$Q^2=0.6$ GeV$^2$, $W=1.95$ GeV]{
\includegraphics[scale=0.35]{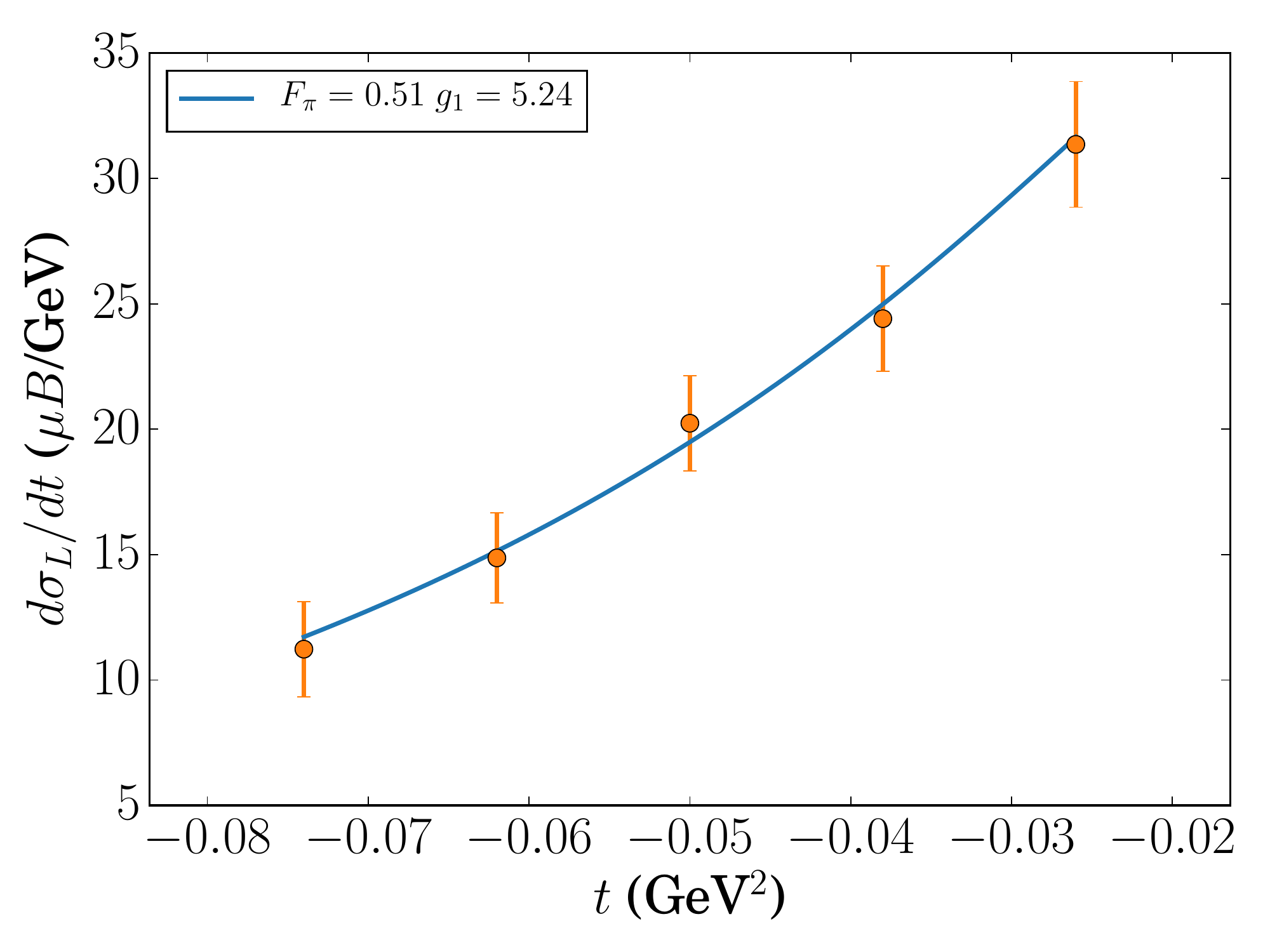}
}
\subfloat[$Q^2=0.75$ GeV$^2$, $W=1.95$ GeV]{
\includegraphics[scale=0.35]{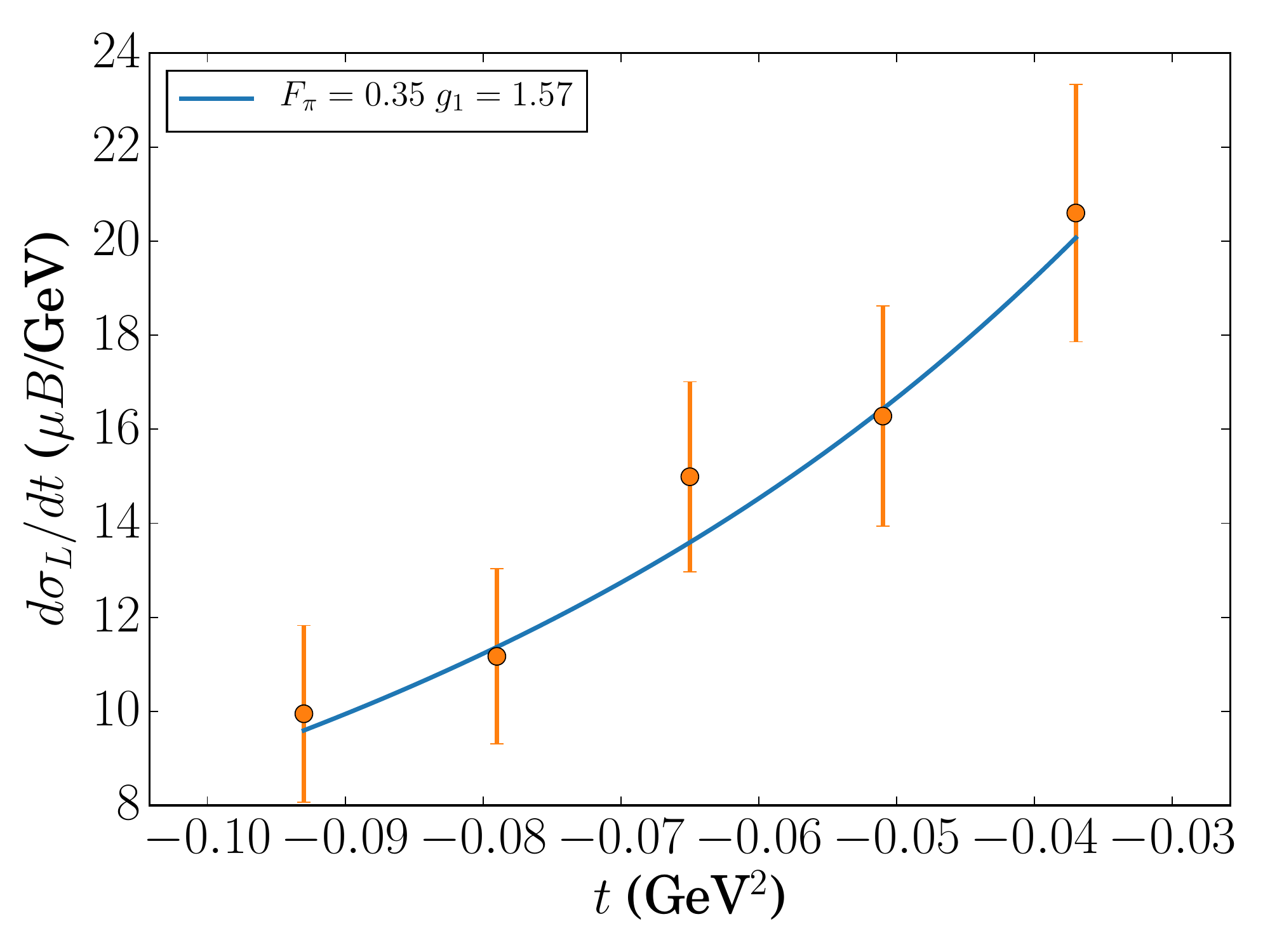}
}

\subfloat[$Q^2=1.0$ GeV$^2$, $W=1.95$ GeV]{
\includegraphics[scale=0.35]{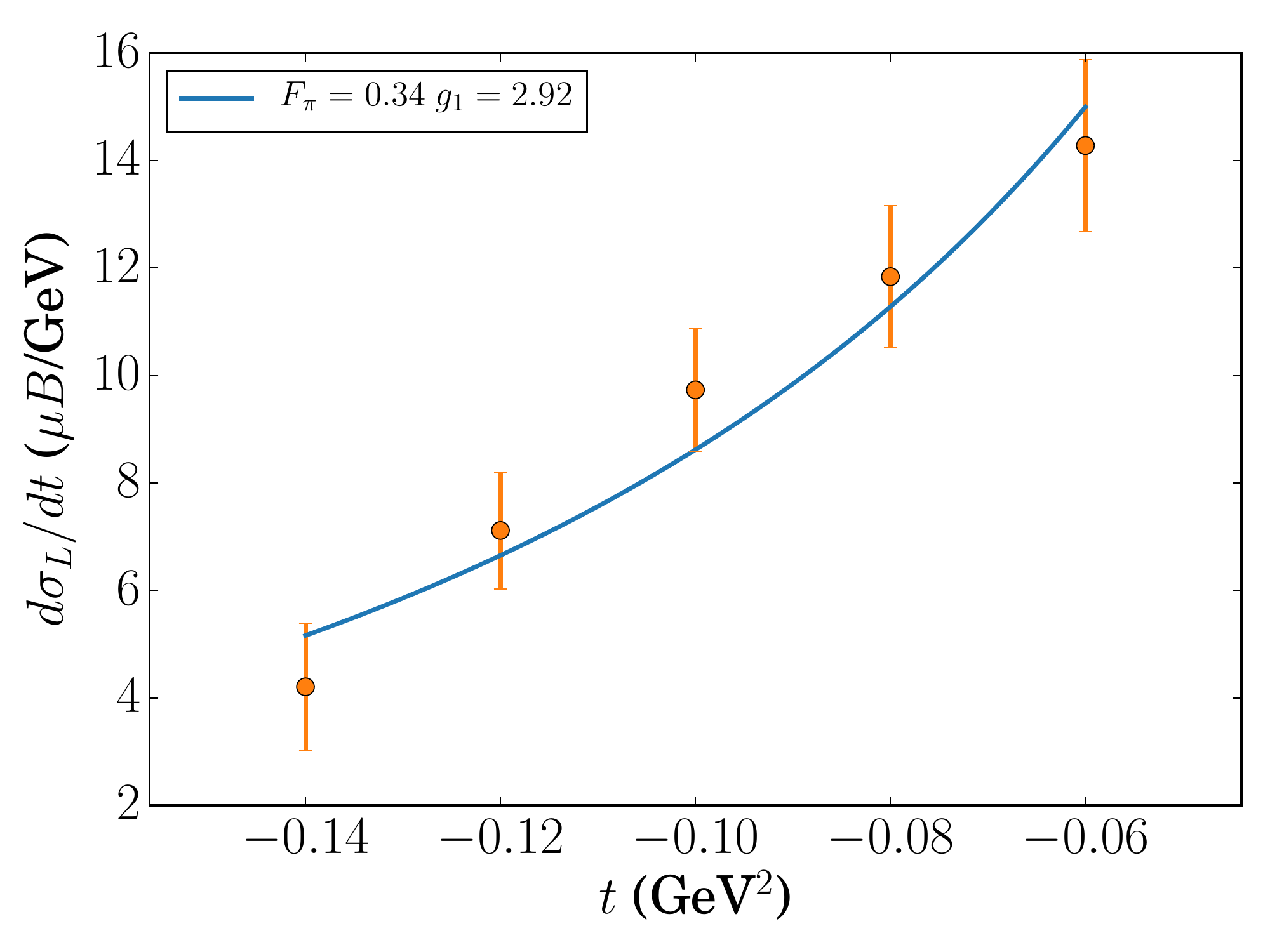}
}
\subfloat[$Q^2=1.6$ GeV$^2$, $W=1.95$ GeV]{
\includegraphics[scale=0.35]{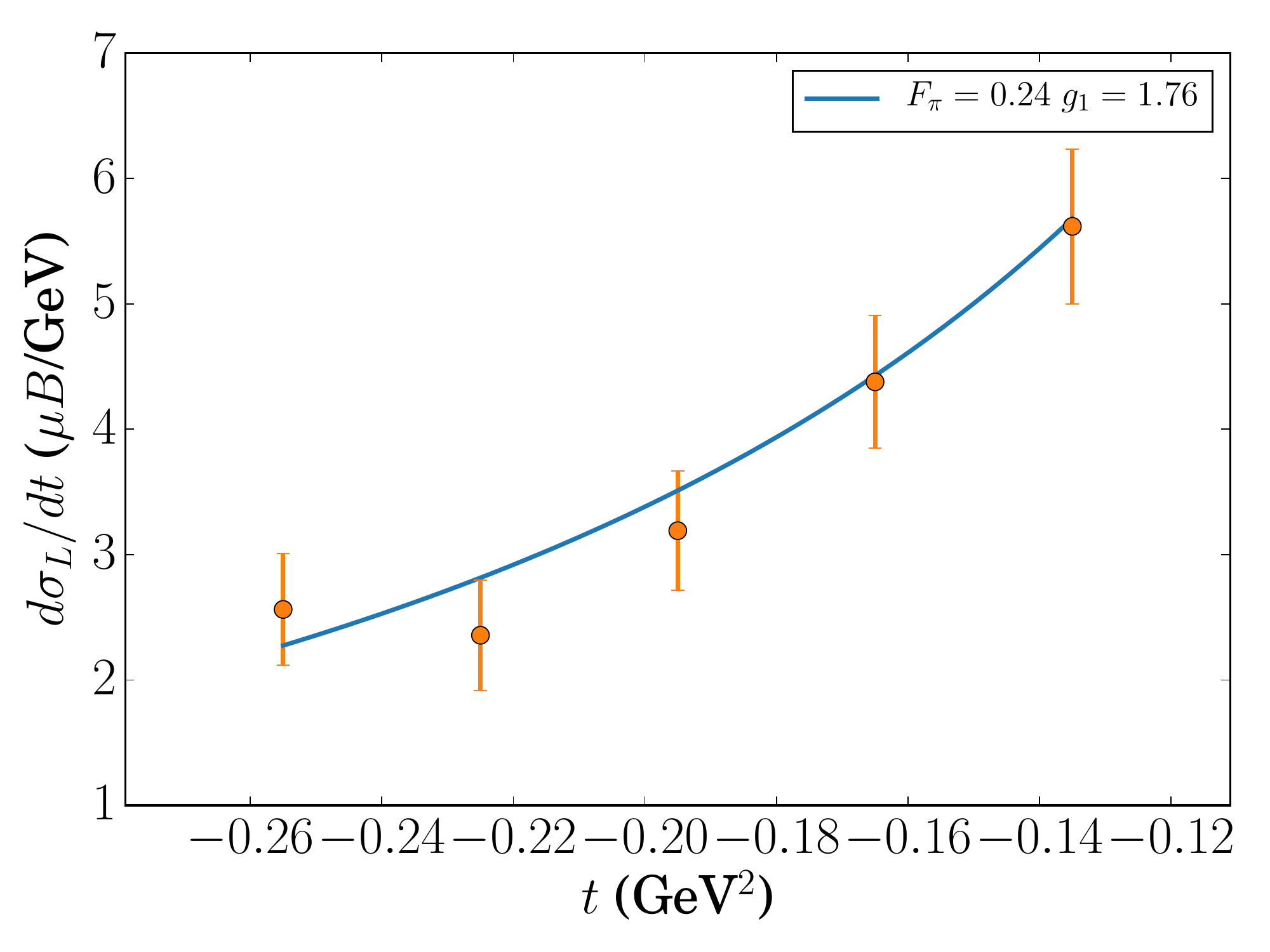}
}

\subfloat[$Q^2=1.6$ GeV$^2$, $W=2.22$ GeV]{
\includegraphics[scale=0.35]{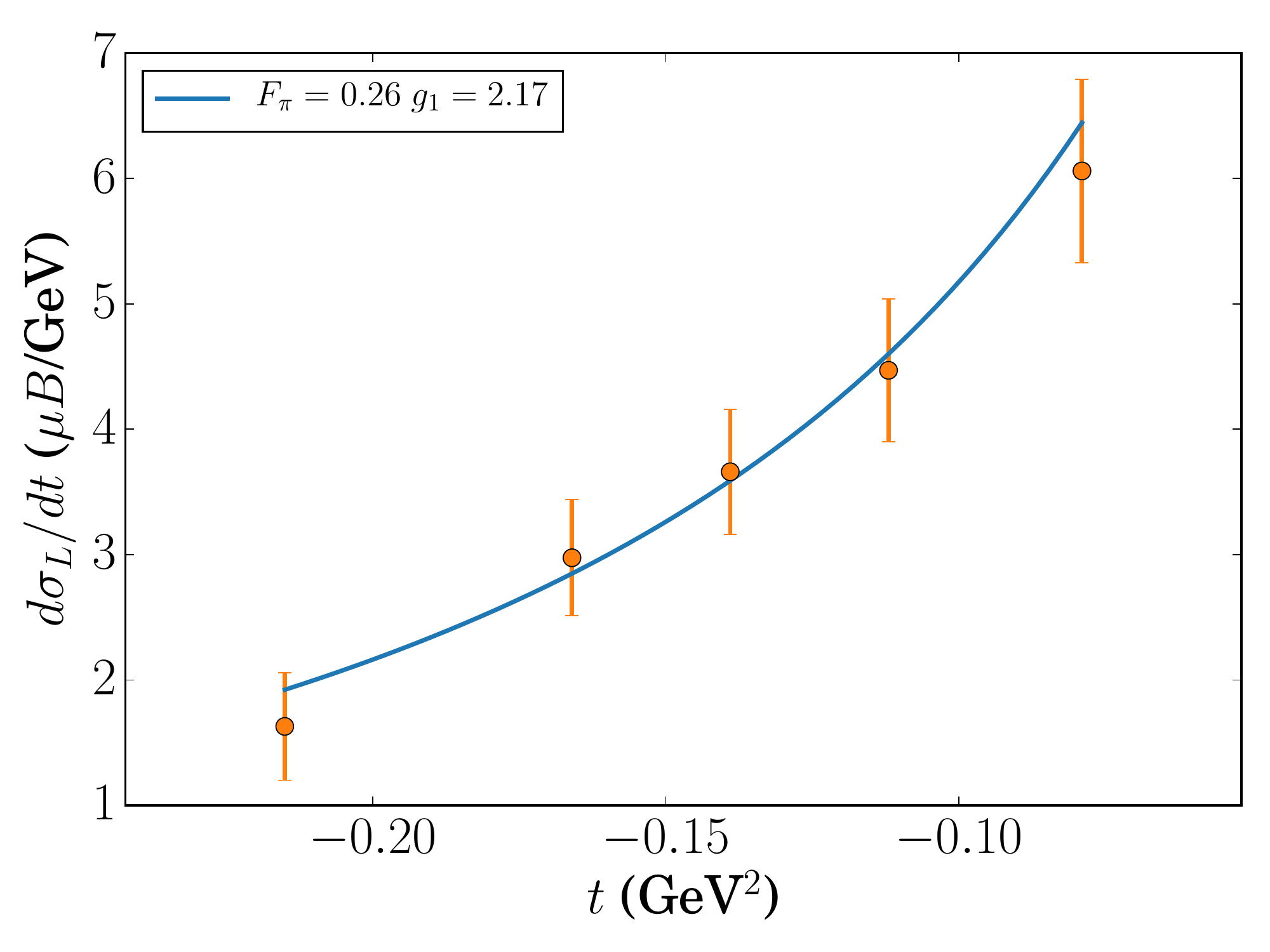}
}
\subfloat[$Q^2=2.45$ GeV$^2$, $W=2.22$ GeV]{
\includegraphics[scale=0.35]{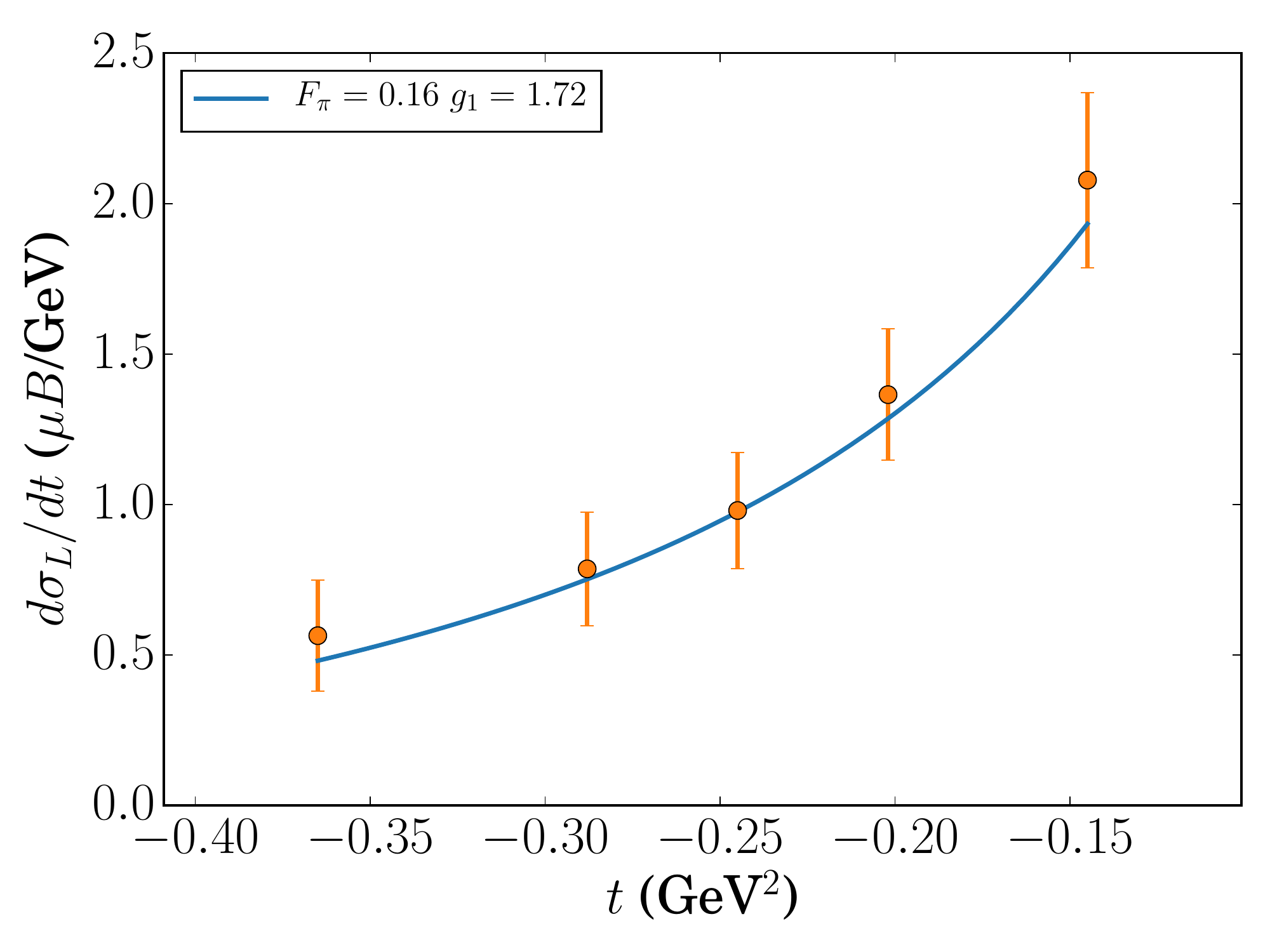}
}
\caption{Fits to the experimental cross section data from Ref.~\cite{Blok:2008jy} using the GI-VGL model.}\label{fig:crossSection}
\end{figure}

\begin{figure}
\includegraphics[scale=0.4]{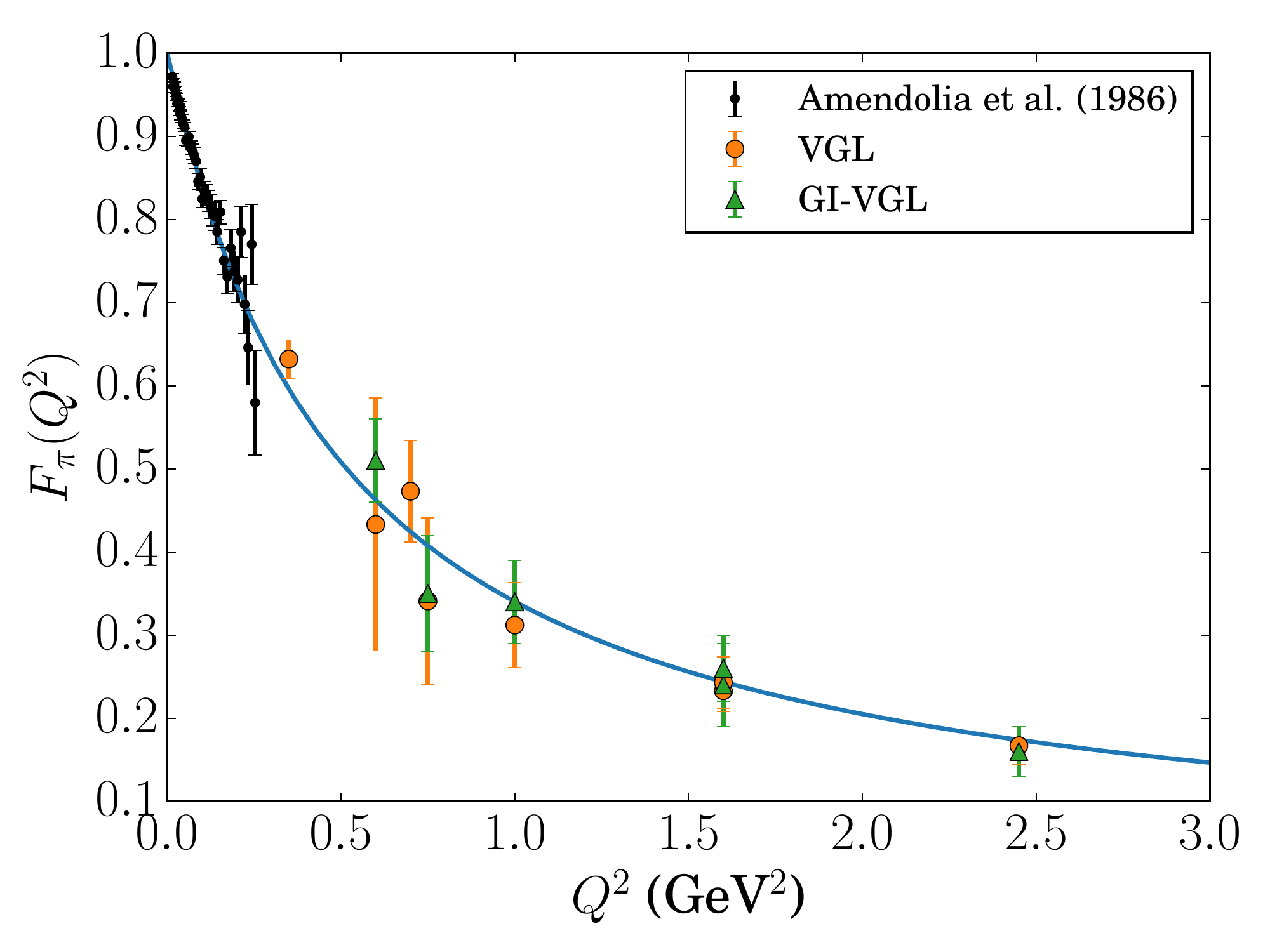}
\caption{Comparison of extracted values of the pion's electromagnetic form factor using the Gauge Improved VGL Model and the previous extraction using the VGL Model. The curve is a fit to the low energy elastic electron pion scattering from Ref.~\cite{Amendolia:1986wj}.}\label{fig:pionFormFactor}
\end{figure}

The resulting fits to the experimental data are given in Fig.~\ref{fig:crossSection}. In order to estimate the uncertainty, we perform scans in $(F_\pi,g_1)$ parameter space for fixed $Q^2$ and $W$, and extract the one sigma contours from which we obtain an estimate of the uncertainties in the model parameters. The resulting extracted pion form factor values with uncertainties are shown in Fig.~\ref{fig:pionFormFactor}, and their exact numerical values are given in Table~\ref{tab:pionFormFactorData}. For comparison, the values obtained from the previous extraction using the VGL Model are also shown. \footnote{Note that the $\chi^2/dof$ improved significantly from the VGL (2.3) to the GI-VGL model (0.5).}

\begin{figure}
\includegraphics[scale=0.4]{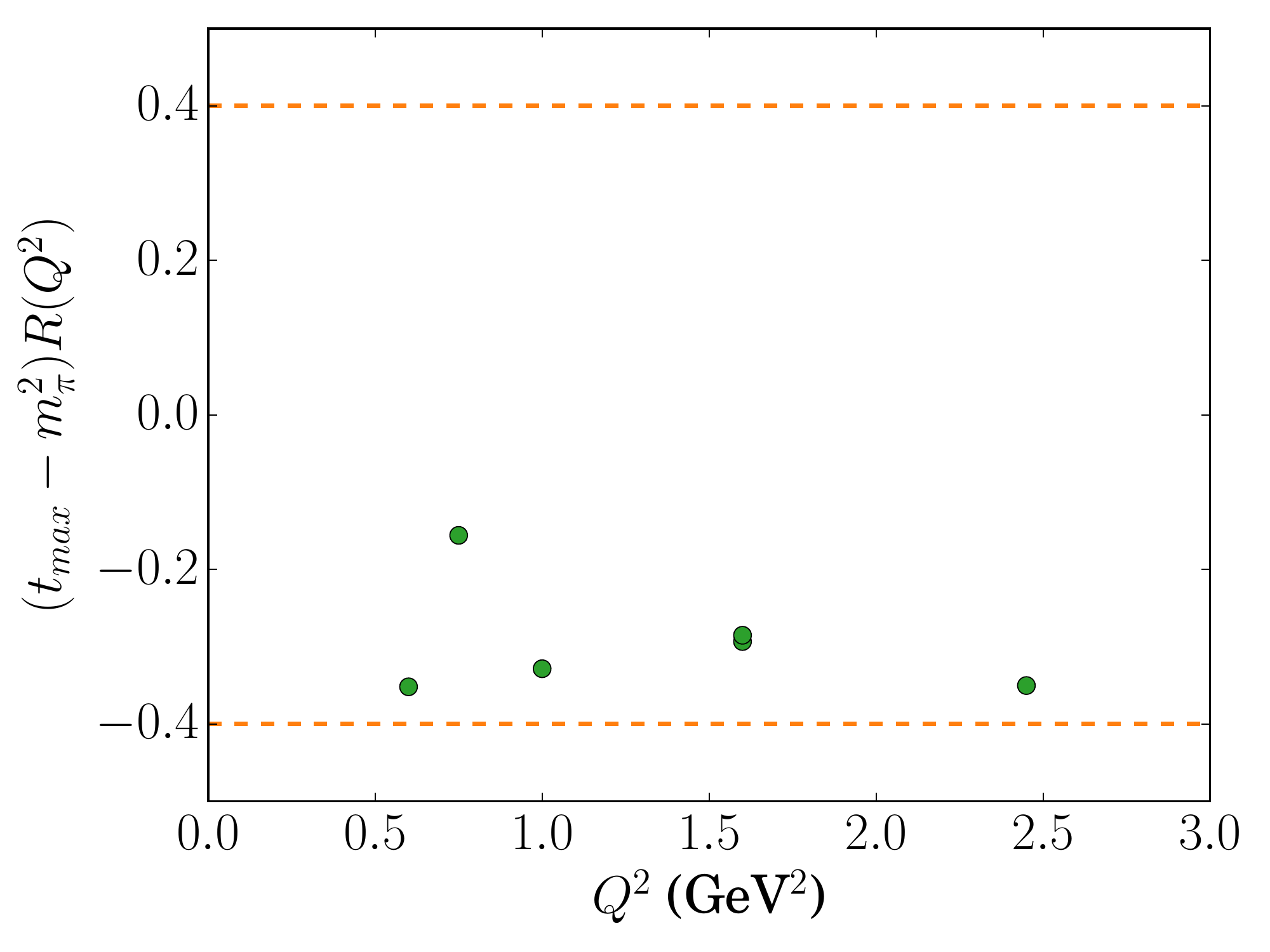}
\caption{Comparison of the magnitude of the first term in the Taylor series expansion of $f_1(t,m_\pi^2,q^2)$ compared to the size of the zeroth term. Symmetric lines at $(t-m_\pi^2)R(q^2)\pm 0.4$ have been added to guide the eye.}\label{fig:ratio_Data}
\end{figure}

When examining the extracted values of the pion form factor, we see that the two models give similar values for all data points considered. Indeed, the two models agree on the extracted pion form factor values within errors. Fig.~\ref{fig:ratio_Data} again compares the contribution from the pole term and the contact term. We note that the trend we observed in the toy model towards larger corrections from the contact term at higher $Q^2$ does not appear to be present here. Instead, the ratio appears to be flat, perhaps implying that the variation in the pion's form factor away from the on-shell point is slower in reality than in our simple toy model. These two points together appear to suggest that the implementation of gauge invariance in the VGL Model does not lead to any major systematic deviation of the extracted pion form factor from the physical value. Indeed, the results reaffirm the reported measurements of the pion form factor.

\onecolumngrid

\twocolumngrid

\section{Conclusion}

The pion's electromagnetic form factor is a subject of continued experimental and theoretical interest because of the theoretical predictions which appear to be testable with the current era experimental facilities.  However, a successful measurement depends on the reliability of the hadronic model used to extract the pion form factor. 

In this paper, we proposed a modification to the existing hadronic model, the VGL Model, which we term the Gauge Improved VGL Model. We produce this model by describing an alternative implementation of gauge invariance, which allows us to use different form factors at the nucleon and pion electromagnetic vertices. As a result of this different method for implementing gauge invariance, we obtain an independent extraction of the pion form factor from electroproduction data. In the case of the experimental data, we found results consistent with the previous extraction. Since our approach removes the requirement that the nucleon and pion electromagnetic form factors are equal, this gives further confidence in the robustness of the measured values of the pion form factor, and in particular that the approach to ensuring gauge invariance does not lead to noticeable systematic errors in the extraction.

\begin{table}
\caption{Comparison of extracted values of pion form factor. Note that we have only included the statistical uncertainty, while the previous has both statistical and systematic uncertainties.}\label{tab:pionFormFactorData}
{\renewcommand{\arraystretch}{1.5}
\begin{ruledtabular}
\begin{tabular}{c c c c}
$Q^2$ (GeV$^2$) & W (GeV) & $F_\pi$ (This analysis) & $F_\pi$ (Ref.~\cite{Huber:2008id})\\ \hline
0.60&1.95&$0.51\pm0.05$& $0.433 \pm 0.017_{-0.036}^{+0.137}$ \\
0.75&1.95&$0.35\pm0.07$& $0.341 \pm 0.022_{-0.031}^{+0.078}$\\
1.00&1.95&$0.34\pm0.05$& $0.312 \pm 0.016_{-0.019}^{+0.035}$\\
1.60&1.95&$0.24\pm0.05$& $0.233 \pm 0.014_{-0.010}^{+0.013}$\\
1.60&2.22&$0.26\pm0.04$& $0.243 \pm 0.012_{-0.008}^{+0.019}$\\
2.45&2.22&$0.16\pm0.03$& $0.167 \pm 0.010_{-0.007}^{0.013}$\\
\end{tabular}
\end{ruledtabular}
}
\end{table}

\begin{acknowledgments}
We thank R. Young and A. Scaffidi for helpful discussions. This work was supported by the University of Adelaide and by the Australian Research Council through the ARC Centre of Excellence for Particle Physics at the Terascale (CE110001104) and Discovery Projects DP150103101 and DP180100497. RJP was further supported by the Taiwanese MoST Grant No. 105-2628-M-009-003-MY4.

\end{acknowledgments}

\newpage

\bibliography{Bibliography}

%
%
%
%
%
%

\end{document}